\documentclass[showpacs,amsmath,amssymb,floatfix,prd,superscriptaddress,footinbib,twocolumn]{revtex4-1}
\usepackage{amssymb}
\usepackage{graphics}
\usepackage{epstopdf}
\usepackage{epsfig}
\usepackage{dcolumn}
\usepackage{bm}
 
\newcommand{\abs}[1]{\left| #1 \right|} 
\newcommand{\avg}[1]{\left< #1 \right>} 
\newcommand{\pd}[2]{\frac{\partial #1}{\partial #2}} 
 
\let\baraccent=\= 
\renewcommand{\=}[1]{\stackrel{#1}{=}} 
\usepackage{color}
 \definecolor{blue}{rgb}{0,0,1}
 \definecolor{sepia}{rgb}{0,0.8,0.2}

\begin{document}

\title{ Finite temperature crossover from a crystalline to a cluster phase for a confined finite chain of ions}



\author{A.V. Zampetaki}
\affiliation{Zentrum f\"{u}r Optische Quantentechnologien, Universit\"{a}t Hamburg, Luruper Chaussee 149, 22761 Hamburg, Germany}
\author{F.K. Diakonos}
\affiliation{Department of Physics, University of Athens, GR-15874 Athens, Greece}
\author{P. Schmelcher}
\affiliation{Zentrum f\"{u}r Optische Quantentechnologien, Universit\"{a}t Hamburg, Luruper Chaussee 149, 22761 Hamburg, Germany}
\affiliation{The Hamburg Centre for Ultrafast Imaging, Luruper Chaussee 149, 22761 Hamburg, Germany}

\date{\today}

\begin{abstract}

Employing Monte-Carlo simulation techniques we investigate the statistical properties of equally charged particles 
confined in a one-dimensional box trap  and detect a crossover from a  crystalline to a cluster phase with increasing temperature. The corresponding transition temperature depends
separately on the number of particles $N$ and the box size $L$, implying non-extensivity due to the long-range character of the interactions.
The probability density of the spacing between the particles exhibits at low temperatures an accumulation of discrete peaks with an overall asymmetric shape.
 In the vicinity of  the transition temperature it is of a Gaussian form whereas in the high temperature regime an exponential decay  is observed. 
 The high temperature behaviour shows a cluster phase with a mean cluster size that first increases with the temperature  and then saturates.
 The crossover is clearly identifiable also in the non-linear behaviour of the heat capacity with varying temperature.
 The influence of the trapping potential on the observed results as well as possible experimental realizations are briefly addressed.
\end{abstract}

\maketitle

\begin{center}
 { \bf{I. INTRODUCTION}} 
\end{center}

Within the past decade, we have witnessed enormous progress with respect to the controlled manipulation of ions. This is in particular due to the application of the dynamics of 
ionic systems to spectroscopy \cite{Thomson93,Hermanspahn00}, the implementation of quantum simulations \cite{Johanning2009, Blatt2012} and the realization of quantum information
processors \cite{Cirac95, Kaler2003}. In this context many trapping methods \cite{Kingdon23, Penning36, Paul90, Yu91, Bluemel95} have been suggested. 
Among the most widely used is the  Paul trap \cite{Paul90} which allows to monitor single ions \cite{Neuhauser80} and study the stability of many ion 
crystals \cite{Drewsen98,Drewsen2002,Drewsen2003}. Experimental studies of ions confined in this type of trap have shown a transition from a cloud of ions to a crystalline structure
 with decreasing  temperature \cite{Diedrich87,Bluemel88}.

Along with the experiments there have also been various theoretical investigations in the formation of Coulomb crystals in the presence of a trapping potential. An early study 
of the structure of spherical Coulomb crystals \cite{Hasse91} showed that particles are arranged in concentric spherical shells with constant inter-shell distances and a hexagonal
surface structure. Furthermore, for large systems a bcc lattice is formed in the interior \cite{Totsuji02} resembling the case of infinite Coulomb systems (OCP) \cite{Ichimaru82}.
 The two-dimensional case has revealed even more exotic phenomena. Specifically, for the cylindrically confined Coulomb lattice a structural phase transition with increasing linear
 density has been found \cite{Hasse90}. The finite 2-D systems of charged particles confined in a parabolic potential or a box presents an order-disorder phase transition
 with increasing temperature \cite{Bedanov94}. The density of the particles differs for the two potentials being almost constant in the inner region and decreasing while moving 
outwards for the parabolic case, whereas it increases radially forming distinct shells for the box. 

Recently, particular interest was dedicated to the study of one-dimensional systems and especially cold ions confined in a corresponding harmonic 
potential \cite{Dubin97, Schiffer03, Morigi04, Morigi2004, Fishman08, Gong10}.
 Analytical approaches have been developed using perturbation theory around the classically minimum energy positions \cite{Morigi04, Morigi2004}. An interesting thermodynamic behaviour different
 from both the 2-D and 3-D case has been observed due to the interplay between the long-range interactions and strong correlations. These observations include a deviation of
 thermodynamic quantities from extensivity, a non-uniform charge density and a structural phase transition (linear to zig-zag)  driven by the strength of the radial potential
 \cite{Fishman08} and temperature \cite{Gong10}.

Following the direction of the above studies, the present paper aims at describing classically the thermodynamic behaviour of equally charged particles confined to an one-dimensional
 box within a wide temperature range. Such a problem is usually treated by computing the partition function of the system:
\[Z(L,N,T) = Z_{K.E} Z_{U},\]
 \begin{equation}
Z_{U}= \int_{0}^{L} \ldots \int_{0}^{L} \exp\left(-\beta V_C(x_1,\ldots x_N) \right) dx_1\ldots dx_N, \label{eq:partg}
\end{equation}
where $Z_{K.E}$ is the part of the partition function due to the kinetic energy of the particles, $\beta=1/ k_B T$, $g^2$ the coupling constant and
\begin{equation} 
V_C(x_1, x_2,\ldots x_N)=\frac{1}{2}{\sum_{\substack{i=1\\   i \neq j}}^{N}}\sum_{j=1}^{N} \frac{g^2}{\abs{x_i-x_j}}
\label{eq:coulomb}
\end{equation}
the Coulomb potential energy. Since $Z_U$ is not analytically accessible, we will proceed here by utilizing  Monte-Carlo simulation techniques. \\
Examining  the  distributions of the spacing between the particles  with increasing temperature, we will observe a transition from a discrete overall asymmetric 
form to a continuous exponential one. This fact can be interpreted as an evidence for a crossover from a crystalline to cluster phase at a transition temperature $T_c$, at which 
this distribution acquires a symmetric Gaussian form. The probability of forming clusters of increasing size increases with the temperature and finally saturates.
We will then proceed and verify the crossover   by the temperature dependence of the heat capacity. Our results indicate that the transition temperature $T_c$ and consequently 
every thermodynamic quantity depends on 
the number of particles $N$ and the  length $L$ of the box in an independent manner, thus presenting deviations from extensivity. 
 An implementation of this non-extensivity in order to obtain finite quantities in the thermodynamic limit is attempted. We also discuss 
the effect of the trapping potential on the observed results.

The paper is organized as follows. In Sec. II we present, in some detail, the Monte-Carlo methods used in our simulations. Sec. III contains our results and a discussion of
 the observables used to characterize the statistical mechanics of the Coulomb system.   In Sec. IV we analyze
 the changes of the observed properties when trapping conditions are altered, 
focusing on the case of harmonic trapping.  Finally,  Sec. V provides a brief summary of our findings commenting on a possible
 experimental realization.

\begin{center}
 { \bf{II. MONTE-CARLO APPROACHES TO THE FINITE ION CHAIN}} 

\end{center}

In order to explore the statistical properties of the finite Coulomb chain confined in a box we employ the Metropolis and Wang-Landau algorithms, each possessing a better efficiency
with reference to the computation of different quantities  in different temperature regimes. In order to be self-contained, we will provide below a brief description of them,
 discussing their advantages and deficiencies and explaining how they are employed in this work. Before doing so, it is important to notice that for a given number $N$ of particles
 in the box the statistical properties of the Coulomb chain are determined by a single dimensionless parameter $\Lambda=\frac{g^2}{L k_B T}$. This is seen by introducing the dimensionless variables $\xi_i=\frac{x_i}{L}$ in eq.~(\ref{eq:coulomb}) and inserting the resulting expression in the partition function (\ref{eq:partg}). In the following we will exclusively use the dimensionless position variables $\xi_i$.
 As a consequence the length of the box can be taken without loss of generality as $L=1$.

For the Metropolis algorithm \cite{Metropolis53} we generate initially a random configuration $\{\xi_i\}$ of the location of the particles in the box. 
The configuration is sorted in an ascending order ($0 \leq \xi_1\leq \xi_2 \leq \ldots \leq \xi_N \leq 1$). Then we choose for the j-th particle a new position 
$0 \leq \xi'_j \leq 1$. The efficiency of the algorithm is greater if we impose the additional restriction $\xi_{j-1} \leq \xi'_j \leq \xi_{j+1}$.
 If  $\Delta E = V_C ({\xi'_i})-V_C ({\xi_i}) \leq 0$ the new configuration is accepted, i.e. $\xi_j=\xi'_j$. Otherwise we accept the 
new configuration with a probability $P$ given by the Boltzmann factor $P=e^{-\beta \Delta E}$. This procedure defines a Monte-Carlo step (MCs) and after a considerable number of 
repetitions the equilibrium state is reached. The Metropolis algorithm satisfies the property of detailed balance and therefore it always converges, being also as a Markov chain
 very efficient in the evaluation of the equilibrium configuration of the particles.   However, it presents also some well-known deficiencies. In the 
low temperature regime, where the minimum of the potential dominates the statistical properties of the system, the acceptance ratio of Metropolis becomes very low, a fact
that leads to a dramatic growth of the simulation time. The situation is even worse when the potential possesses many minima, since the particles can be trapped in one of
 them and never reach the global minimum. For the strongly correlated system presented here, the main problem arises in the calculation  of inherently averaged macroscopic quantities
 like the heat capacity $C_V$.  A large ensemble of configurations is needed in order to reduce statistical errors significantly and thus the practical simulations are prohibited.

Recently another algorithm has been proposed in order to overcome these problems, namely the Wang-Landau algorithm \cite{Wang2001}. Contrary to the canonical ensemble based
 Metropolis, this algorithm uses the concept of the microcanonical  ensemble and it intends to calculate the density of states (DOS) of a system as described below.
First of all we choose the range of accessible energies and then divide it into a number of bins $M$. We start, assuming that we have a uniform density of states $g(E)$ 
i.e. $g(E)=1$ for every energy bin. We then proceed as follows. For the j-th particle we choose a new position $\xi_{j-1} \leq \xi'_j \leq \xi_{j+1}$. 
If $g(E_2) \leq g(E_1)$ with $E_1=V_C ({\xi_i})$ and $E_2=V_C ({\xi'_i})$ we accept the new configuration. Otherwise the new configuration is accepted with a probability 
$P=\frac{g(E_1)}{g(E_2)}$. Each time an energy bin is visited we update the corresponding density of states by multiplying the existing value by a modification factor $f$, 
i.e. $g(E)=f \cdot g(E)$. We choose $f=e^1 \approx 2.71828..$. We also update the energy histogram $H(E)=H(E)+1$. In the original version of the algorithm \cite{Wang2001} for 
discrete systems, the above steps were repeated until a flat histogram was obtained (e.g $\min(H(E))  \leq 0.8 \cdot \avg{H(E)} $, with $\avg{H(E)}$ being the mean value of the 
histogram). In order to improve the accuracy which is of order $\ln f$ we decrease the modification factor $f=\sqrt{f}$ and repeat the procedure. When $\ln f \approx 10^{-8}$ the 
density of states does not any longer  evolve any further and the simulation is stopped. The major problem of this algorithm is that it does not satisfy the detailed balance 
condition and so its  convergence can not be strictly proved. Furthermore the method has been suggested originally for discrete systems with narrow energy landscapes. 
Nevertheless it has been used successfully in recent calculations considering also continuous systems \cite{Panag02,Pablo02, Poulain06}. 
If the low energy spectrum of the system is complex the criterion of the flatness of the histogram is never satisfied as some energy bins are never visited.
 To overcome this problem several alternatives have been proposed \cite{Dellago05, Poulain06}. In our work we  use the technique described in \cite{Dellago05} 
 excluding some boundary bins from the flatness check. The Wang-Landau algorithm has the special advantage  
that once the DOS is obtained, we can easily derive all the thermodynamic quantities (e.g. $Z=\sum_{E} g(E) e^{\beta E}$) from this  function alone. 
Thus, the simulation time is significantly reduced when computing the temperature dependence of demanding quantities like the heat capacity $C_V$.  However, for 
computing properties that depend on the position of the particles for systems with a complex, degenerate and unbounded energy landscape as the one presented here, this method is 
in an inferior position in comparison to Metropolis. This is attributed to the large computational effort  needed in order to find  a representative sample of microstates for each
 energy $E$. In order to exploit the advantages of each algorithm avoiding its drawbacks, we use both of them and apply each one for the evaluation of different properties.
 
In the present study $1.5 \cdot 10^7$ MCs were required in order to reach equilibrium with the Metropolis algorithm and an ensemble of $200$ configurations in order to obtain
 the mean energy $\avg{E}$ as a function of $\Lambda$ and the number of particles $N$. For quantities related to the positions of the particles at equilibrium the size of the 
ensemble was increased up to $10000$ configurations to achieve better statistics. The Wang- Landau algorithm has been realized with $2 \cdot 10^7$ MCs for each value of the 
modification factor until $\ln f \approx 10^{-8}$ as stated above. In most calculations the number of particles $N=100$ is used unless stated otherwise.

The spatially resolved density and correlation functions, as well as  $\avg{E(N)}$  have been computed with the Metropolis whereas the heat capacity $C_L(\Lambda)$ 
has been determined using the Wang-Landau algorithm. 
As a check, the dependence of the mean potential energy $\avg{E(\Lambda)}$ on the parameter $\Lambda$  has been computed with both approaches.
 
\begin{figure}[htbp]
\begin{center}
\includegraphics[width=8.6cm]{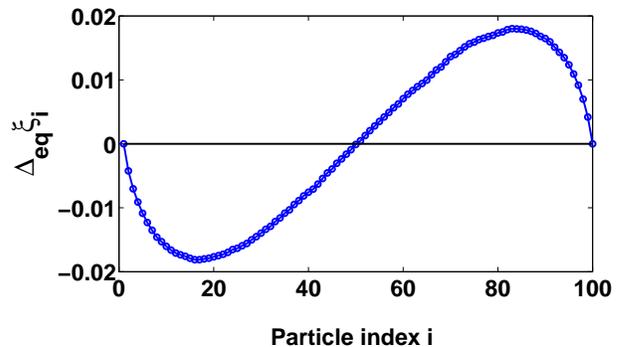}
\end{center}
\caption{\label{deltaeq1} (color online). The difference of the expected equidistant position of each particle from its minimum energy position  
as a function of the index of the particle for $N=100$ particles confined in a box with unit length $L=1$.}
\end{figure}

\begin{figure}[htbp]
\begin{center}
\includegraphics[width=8.6cm]{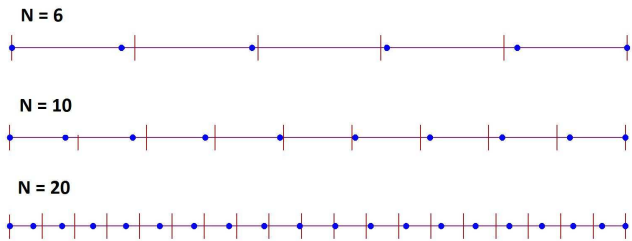}
\end{center}
\caption{\label{conf1} (color online). The minimum energy configurations of particles as obtained by the minimization of the potential $V_C (x_1, x_2, \ldots, x_N)$ for 
$N=6, 10$ and $20$. The particles are represented with the (blue) full dots and the equidistant positions are marked with the (red) vertical line segments.}
\end{figure}

\begin{figure}[htbp]
\begin{center}
\includegraphics[width=8.6cm]{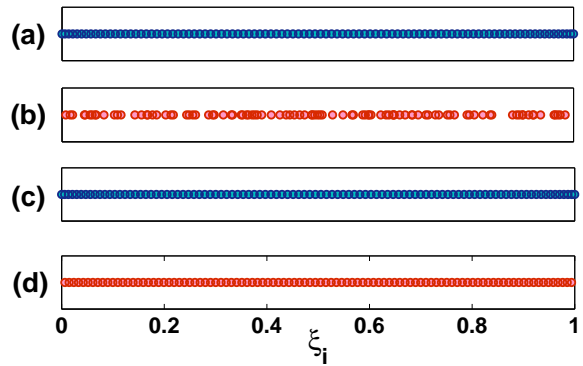}
\end{center}
\caption{\label{conf2} (color online). (a-d) Particle configurations for $N=100$: (a) A single configuration of particles at $\tau=10^{-3}$,
 (b) Same for $\tau=10^3$, (c) The mean configuration of particles for $\tau=10^{-3}$, 
(d) Same for $\tau=10^3$.}
\end{figure}

\begin{center}
 { \bf{III. NUMERICAL RESULTS AND DISCUSSION}} 
\end{center}
We present and describe in the following our basic results for the temperature-dependent behaviour and properties of the ions in the box. Let us begin with the exploration of the
 minimum energy configuration (Subsec. A) and continue with its temperature dependence (Subsec. B).
 In the subsection C which addresses the main results of this work evidence for the crystalline-clustering crossover and the associated phenomenology is presented. 
\begin{center}
 { \bf{A. Minimum energy  configuration}} 
\end{center}
Some useful remarks on the minimum energy configuration of the ions in the box can be obtained simply by inspection of the corresponding Coulomb potential energy. First of all 
$V_C(\xi_1,\xi_2,\ldots, \xi_N)$ is obviously a strictly increasing function of $\xi_1$ ($\pd{V}{\xi_1}>0$)  and a strictly decreasing function of $\xi_N$ ($\pd{V}{\xi_N}<0$).
Thus, the first and the last particles always occupy the edges of the line segment i.e. $\xi_1=0$ and $\xi_N=1$. It is obvious that the problem possesses a symmetry axis with 
respect to the centre of the box. As a result, if the number of particles $N$ is odd, the  central particle is positioned at the centre of the line segment 
($\xi_{\left\lfloor N/2\right\rfloor +1}=1/2$, where $\left\lfloor x \right\rfloor$ denotes the floor function of the number $x$.). Furthermore, for more than $3$ particles the  minimum energy configuration is not the equidistant one. In fact, the difference $\Delta_{eq} \xi_i=\xi^0_i -\xi^{eq}_i$ with  $\{\xi^{0}_i\}$ the equilibrium positions of the particles at 
zero temperature and $\{\xi^{eq}_i=\frac{i}{N-1}\}$
the equidistant positions,  is a discretization of a smooth function, 
inversion symmetric w.r.t. the centre position $N/2$ possessing a minimal value at $i_{min}\lesssim N/4$, a maximal value at $i_{max}\gtrsim 3N/4$ and being almost linear within 
$[i_{min}, i_{max}]$ (Fig. \ref{deltaeq1}).  
Due to the fixed  length of the box and the long-range character of the interactions, the particles tend to accumulate at the edges of the box leaving larger inter-space distances 
in its middle part (Fig. \ref{conf1}, \ref{conf2} (a)). This is opposite to the case of ions confined in an 1-D harmonic potential \cite{Dubin97, Morigi2004} where 
the length of the chain is not fixed and the particles tend to accumulate in the inner region. Such a behaviour could be expected as the one-dimensional analogue of 
the 2-D system presented in  \cite{Bedanov94}.  In the specific case of the 1-D system with the first and the last particle fixed at the edges, 
it can be proven (see appendix), that due to the ordering of positions, the lowest energy configuration presented here provides the only minimum of the potential energy.
This is in contrast to  systems in higher dimensions, such as 3-D ionic systems under harmonic confinement, where many local minima and multiple isomers exist 
\cite{Calvo07}.  

\newpage
\begin{center}
 { \bf{B. Temperature dependence of the configurations and the densities of charges}} 
\end{center}

To study the temperature dependence of the observables describing the properties of the charged particles in the box we have to vary the parameter $\Lambda$.
 We define $\frac{g^2}{L}=k_B T_0$ yielding $\tau=\frac{1}{\Lambda}=\frac{T}{T_0}$ as a reduced temperature.
As $\tau$ increases positions different from the minimum energy ones become accessible to the particles. At temperatures $\tau \gtrsim 10$ the particles start to
form clusters of different sizes (Fig. \ref{conf2} (b)). Thus the form of the possible particle configurations changes dramatically. However, if we consider the mean positions of 
the particles (Fig. \ref{conf2} (c),(d)) over an ensemble of $10^4$ configurations we observe that for the high temperature case 
(Fig. \ref{conf2} (d)) the clustering is averaged out and the resulting mean configuration  resembles very much that of the low temperature regime (Fig. \ref{conf2} (c)).
 A slight difference, however,  is that for high temperatures the positions seem closer to the equidistant ones with less accumulation at the edges of the box, which are no
 longer occupied.

\begin{figure}[htbp]
\begin{center}
\includegraphics[width=8.6cm]{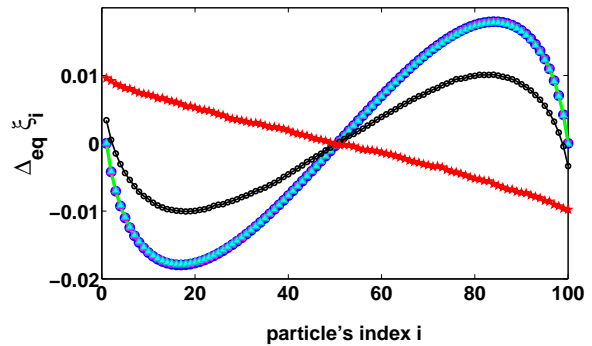}
\end{center}
\caption{\label{deltaeq2} (color online). The mean difference  $\Delta_{eq} \xi_{i}$ of the position of each particle from its expected equidistant position as a function of the 
particle's index $i$  for  temperatures $\tau=10^{-5}$ (blue circles), $10^{-3}$ (green line), $10^{-1}$ (magenta squares), $1$ (cyan triangles), $10^2$ (black line with circles)
 and $10^4$ (red stars). Note that the  $\tau=10^{-1}, 1$ curves for $\Delta_{eq} \xi_{i}$ are on top of those for $\tau=10^{-5}, 10^{-3}$ and therefore hardly visible.}
\end{figure}

The properties of the mean equilibrium configuration for different temperatures can be further explored by considering the temperature dependence of the quantity
 $\avg{\Delta_{eq} \xi_i}$ and the density function $\avg{\rho(\xi)}$. As seen in Fig. \ref{deltaeq2} for temperatures $\tau \leq 1$ the mean positions of the particles are
 identical with the minimum energy ones (Fig. \ref{deltaeq1}). 
Even at $\tau=10^2$ (black line with circles) the form of $\avg{\Delta_{eq} \xi_i}$ is preserved except from a decrease in its magnitude and a deviation at the edges. 
However, for $\tau = 10^4$ the form becomes linear keeping the symmetry around the centre but changing by an overall sign.

\begin{figure}[htbp]
\begin{center}
\includegraphics[width=8.6cm]{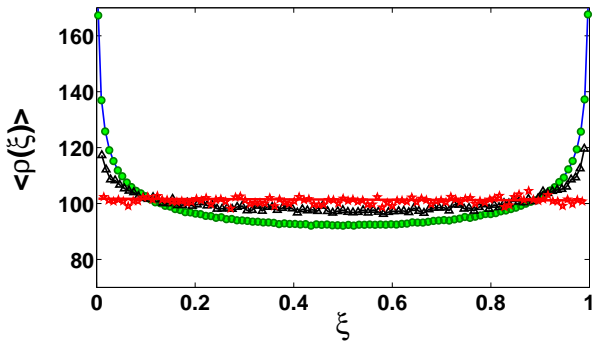}
\end{center}
\caption{\label{dens1} (color online). The density of charge $\avg{\rho(\xi)}$ as a function of the particles' positions $\xi$ for  temperatures $\tau=10^{-5}$ 
(blue line), $1$ (green circles), $10^3$ (black line with triangles), $10^5$ (estimated mean red line and fluctuations red stars). 
The particle number is $N=100$.}
\end{figure}

A reference to the density of the absolute positions ${\xi_i}$ would be of no practical use since, due to the finite number of particles, the problem is discrete and 
it would only result in a set of delta functions.
In previous studies \cite{Dubin97, Morigi04, Morigi2004} there has been a particular interest in the quantity $\Delta \xi_i=\xi_{i+1}-\xi_{i}$  denoting the inter-space distance 
between successive particles. It has been shown that its inverse is a smooth function of $\xi_i$ and represents the density of ions per unit length. To improve statistics we use the
 ensemble average of $\Delta \xi_i$ i.e $\avg{\Delta \xi_i}$. Through interpolation, we produce the density of charge $\avg{\rho(\xi)}$ in the continuum limit.
 Our results are presented in Fig. \ref{dens1} for various temperatures. We clearly observe the non-uniformity at the outer areas (close to the edges) for temperatures $\tau \leq 1$
 (opposite to what has been observed in  \cite{Dubin97}) and an almost uniform behaviour for high temperatures
except for the region close to the edges where the density becomes zero (not visible in Fig. \ref{dens1}). As expected, there exist large thermal fluctuations in the high temperature 
regime.

\begin{figure}[htbp]
\begin{center}
\includegraphics[width=8.6cm]{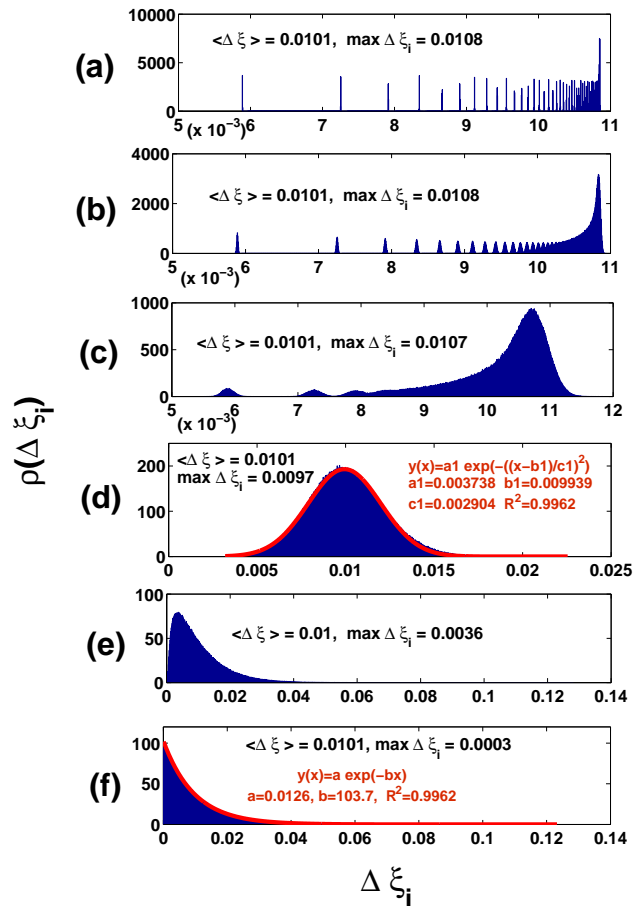}
\end{center}
\caption{\label{difx1} (color online). (a-f) The distributions of $\Delta \xi_i$, $\rho(\Delta \xi_i)$, for  different temperatures: (a) $\tau=10^{-5}$, 
(b) $\tau=10^{-3}$, (c) $\tau=10^{-1}$, (d) $\tau=10$ (the red line is a Gaussian fit), (e) $\tau=10^3$, (f) $\tau=10^5$ (the red line is an exponential fit).
In each figure  the mean value of the inter-space distance $\avg{\Delta \xi}$ and its most probable value ($\max \Delta \xi_i$) are provided.}
\end{figure}

\begin{center}
 { \bf{C. Crossover from the crystalline to cluster phase}} 
\end{center}
As has been stated above the quantity $\Delta \xi_i$ is very useful in the determination of the density of charges $\rho(\xi)$, a continuum limit of the distribution of 
the equilibrium positions $\{\xi_i\}$. In this subsection we present the results for the distribution of $\Delta \xi_i$, $\rho(\Delta \xi_i)$ at various temperatures 
(Fig. \ref{difx1}). It turns out that this quantity contrary to  $\rho(\xi_i)$ is well-defined for this discrete system and even more, it carries significant information for its 
thermodynamic behaviour. It has been computed through the histograms of  $\Delta \xi_i$ for $10^4$ configurations and $N=100$. For low temperatures 
 (Fig. \ref{difx1} (a), (b)) a crystalline structure is observed as $\rho(\Delta \xi_i)$ contains clearly a part formed by discrete peaks for small distances. Thus, the particles occupy
 well-defined positions leaving specific distances between each other. The mean value of $\Delta \xi_i$ for all the histograms approximately coincides with the equidistant
 spacing $\avg{\Delta \xi}\approx 0.0101\approx \frac{1}{N-1}$. The most probable value ($\max \Delta \xi_i \approx 0.0108$) is slightly larger than the
 mean, denoting an anti-clustering effect and a crystallization of  ions. The peaks are broadening and merging with an increase of the temperature and the overall distribution becomes
 smoother (Fig. \ref{difx1} (c)). More positions become accessible to the particles and the most probable value decreases tending to the mean one. At $\tau=10$ the form of
 $\rho(\Delta \xi_i)$ becomes almost symmetric (Fig. \ref{difx1} (d)) and its line shape resembles  a Gaussian as is verified by a $\chi^2$-fit (red line). 
Here $\max \Delta \xi_i  \approx 0.0097$ is slightly less but actually very close to $\avg{ \Delta \xi}$. In the high temperature regime we observe a clustering effect 
(Fig. \ref{difx1} (e) ) right opposite to that of the low temperature case. The effect is strongly pronounced as can be seen by inspecting $\max \Delta \xi_i \approx 0.0036$
 which is much less than the mean $\Delta \xi_i$ value. This trend is enhanced for ultra-high temperatures ($\tau=10^5$) as illustrated  in Fig. \ref{difx1}(f). 
The most probable value of the spacing is almost zero and the distribution is well-described by an exponential  
(red line fit with a coefficient of determination $R^2=0.9962$ \cite{everitt}). Its characteristic  length scale is approximately $0.0096$ which is also close to the equidistant
 spacing of the particles $\frac{1}{N-1}$. This reflects the fact that for high temperatures the particles have the opportunity to occupy almost every available position and 
each particle's position is completely independent from the positions of the neighboring particles.
The continuous change of $\rho(\Delta \xi_i)$ implying the transition from a crystalline  to a cluster configuration with the change of temperature is a strong indicator
 of a crossover occurring at a transition temperature $\tau_c \approx 10$, where the distribution becomes Gaussian (Fig. \ref{difx1} (d)).

\begin{figure}[htbp]
\begin{center}
\includegraphics[width=8.6cm]{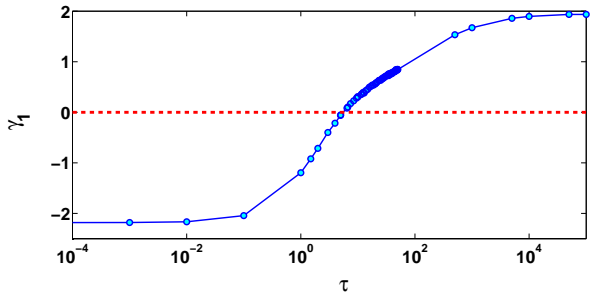}
\end{center}
\caption{\label{skew1} (color online).  The skewness of the distributions  $\rho(\Delta \xi_i)$ as a function of the temperature $\tau$. }
\end{figure}

  \textit{Skewness}. A more precise determination of the transition temperature can be achieved by  examining the skewness $\gamma_1$ of the 
distributions $\rho(\Delta \xi_i)$ as a function of the temperature (Fig. \ref{skew1}). Skewness is a measure of the asymmetry of a probability distribution $\rho(x)$ and
is defined as the third standardized moment:
\[\gamma_1=\frac{\avg{x-\avg{x}}^3}{\sigma^3},\]
where $\sigma$ is the standard deviation of $\rho(x)$. It is evident that in our case the skewness goes from negative to positive
 values as temperature increases (Fig. \ref{skew1}). Its absolute value
 possesses a minimum at a temperature $\tau_1\approx 6$, which can be conceived as the critical one. At higher temperatures it saturates at 
the value $2$ which is indeed the skewness of the exponential distribution. 

\begin{figure}[htbp]
\begin{center}
\includegraphics[width=8.6cm]{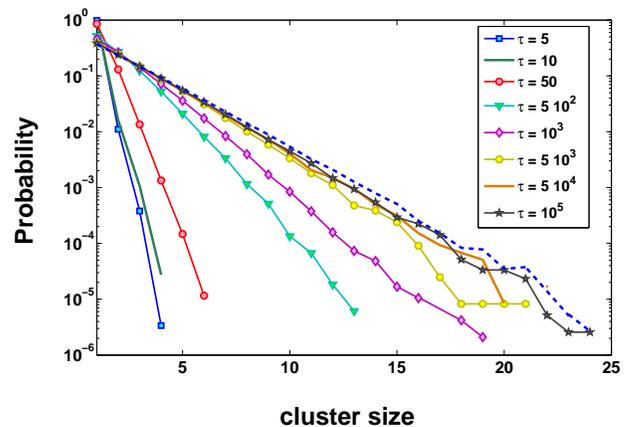}
\end{center}
\caption{\label{clust1} (color online).  The probability distribution of cluster sizes  for  different  temperatures (semi-logarithmic scale): $\tau=5$ (blue line with squares), 
$\tau=10$ (green line),
$\tau=50$ (red line with circles), $\tau=5\cdot 10^2$ (cyan line with triangles), $\tau=10^3$ (purple line with diamonds), $\tau=5 \cdot 10^3$ (yellow line with circles),
 $\tau=5 \cdot 10^4$ (brown line), $\tau=10^5$ (black line with stars). The blue dashed line indicates the probability distribution of cluster sizes in the case of non-interacting 
particles confined in the box (purely random configurations).}
\end{figure}

\textit{Cluster size and temperature}. We have mentioned above that for temperatures 
$\tau >10$ the particles tend to form clusters of various sizes. Let us now explore how the temperature affects the size of the clusters formed. 
In what follows we consider as cluster size $1$ the size of a neighborhood 
which includes only 1 particle i.e. the case when no cluster is formed.   Furthermore, a particle $i$ is assumed to belong to the same cluster as
 its previous neighbor $i-1$ if they are 
separated by a distance smaller than a threshold $d$. The value of this threshold is taken to be the minimum inter-particle distance $\Delta \xi_i$ of the mean configuration 
of particles at temperature $\tau$.  This choice helps in focusing on thermal fluctuations thereby avoiding effects emerging from the  specific structure of the crystal.
 Note that this quantity depends as well  on the number of  particles $N$.
 We observe (Fig. \ref{clust1}) that the probability of forming a larger cluster is generally increased with the increase of the temperature tending to a limiting form 
for ultra high temperatures.   This form coincides with the one obtained for randomly chosen configurations of particles confined in a box of length $L=1$, i.e. for the
non interacting case. Therefore, for arbitrarily high temperatures the singularity of the Coulomb potential at small inter-particle distances becomes irrelevant. 

\begin{figure}[htbp]
\begin{center}
\includegraphics[width=8.6cm]{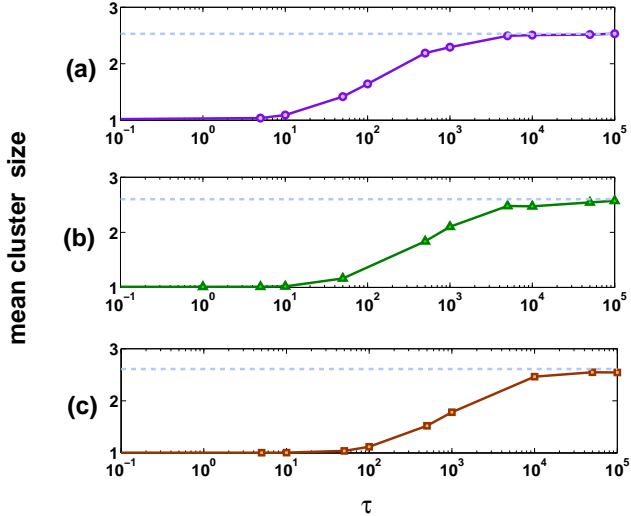}
\end{center}
\caption{\label{clust2} (color online).  (a-c) The mean cluster size as a function of the reduced temperature $\tau$ for systems with different numbers of particles:
(a) N=50, (b) N=100, (c) N=200. The horizontal dashed lines indicate the values of the mean cluster size for the respective numbers $N$ of non-interacting 
particles confined in the box. }
\end{figure}

This behaviour is also reflected in the dependence of the mean cluster size on the temperature which is presented in Fig. \ref{clust2} for the particle numbers  $N=50,100$ and
 $200$.    The variation with the temperature is similar for the three cases being almost constant (one) for low temperatures, then increasing logarithmically 
and finally saturating at a value close to $2.6$, which is essentially the value of the mean cluster size of non-interacting particles confined in the box. 
For $N=100$ the mean cluster size starts to increase around $\tau=10$ which, as we have seen, is close to the 
transition temperature  of the system. For $N=50$ this temperature is slightly less (approximately $\tau=4$) whereas for $N=200$ it  is around $30$. Thus, we can make
the crude statement that the transition temperature for this finite system  increases (faster than linearly) with the number of particles.

\begin{figure}[htbp]
\begin{center}
\includegraphics[width=8.6cm]{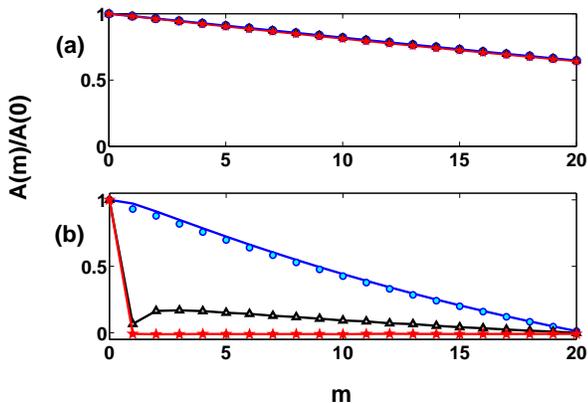}
\end{center}
\caption{\label{auto1} (color online).  (a-b) The normalized correlation functions $A(m)/A(0)$ for $\tau=10^{-5}$ (blue line), $\tau=10^{-1}$ (cyan circles), $\tau=10$ 
(gray triangles with black line), $\tau=10^5$ (red line with stars) and for two different quantities:
(a) the absolute positions of the particles $\xi_i$ (b) the  inter-space distance $\Delta \xi_i$. 
Note that for figure (a) all the curves for different $\tau$ are on top of each other and therefore not distinguishable.}
\end{figure}

\textit{Correlation functions}. Concerning the correlation functions of the particles' positions \[A(m)=\avg{\avg{\xi_{i+m}\xi_{i}}}-\avg{\avg{\xi_{i+m}}}\avg{\avg{\xi_{i}}}\]
 (where $\avg{\avg{\ldots}}$ denotes the average both over the indices $i$ and the ensemble of configurations), we encounter (Fig.~\ref{auto1} (a)) 
a very slow  decay with increasing distance $m$, taken as the magnitude of the difference of the particles' indices.
The form of these functions is very close to a quadratic polynomial. A remarkable feature of the  position correlation 
functions is that temperature does not  affect them contrary to the typical case for most statistical systems (e.g. the spin correlation functions of the classical Ising model \cite{barkema}). This however, is in agreement with the fact that the mean configuration of the particles doesn't change much with the increase of temperature (Figs.~\ref{conf2} (c),(d)).

On the other hand, the correlation functions of the spacing between the particles 
\[A(m)=\avg{\avg{\Delta \xi_{i+m}\Delta \xi_{i}}}-\avg{\avg{\Delta \xi_{i+m}}}\avg{\avg{\Delta \xi_{i}}}\]
 exhibits a  strong dependence on the temperature (Fig.~\ref{auto1} (b)). In particular, for low temperatures 
we observe a slower decay of the correlation function $A(m)$, 
whereas at high temperatures ($\tau \geq 10$) the decay is in general abrupt becoming practically zero for $m \geq 1$.

\begin{figure}[htbp]
\begin{center}
\includegraphics[width=8.6cm]{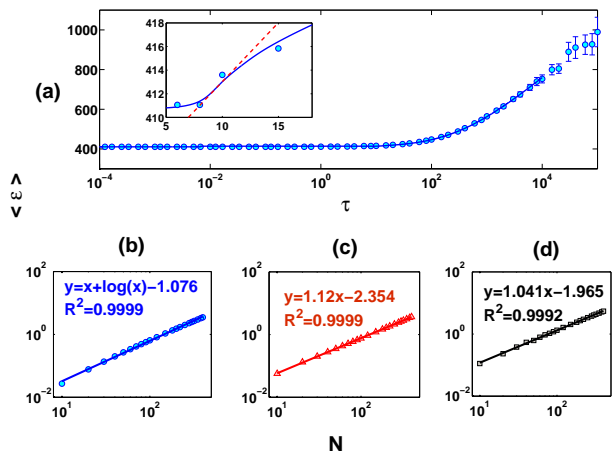}
\end{center}
\caption{\label{mesen} (color online).  (a) The dependence of the dimensionless specific mean energy $\avg{\varepsilon}$ on the reduced temperature $\tau$ on a semi-log scale
for $N=100$ (Metropolis: cyan circles; Wang-Landau: blue line).
(b-d) The dependence of the dimensionless specific mean energy $\avg{\varepsilon}$ on $N$ on a log-log scale for  (b) $\tau=10^{-2}$ (circles for numerical values, blue line  fit),
(c) $\tau=10$ (triangles, red line linear fit), (d) $\tau=10^5$ (squares, black line linear fit).}
\end{figure}

 \textit{Energy dependence on $N$, $T$, $L$}. The mean potential energy is in general a function of
 the temperature $T$, 
the length of the box $L$, and the number of particles $N$ separately i.e. $\avg{E}=f(T, L, N)$.
In order to study numerically this dependence we need to deal with dimensionless quantities. We have already introduced the reduced temperature $\tau$ and the 
dimensionless positions $\{\xi_i\}$. The next step is to introduce the reduced specific mean energy $\avg{\varepsilon}\equiv \frac{L \avg{E}}{g^2 N}$, in analogy with 
the dimensionless reduced temperature $\tau$, and to study the behaviour of $\avg{\varepsilon}=h(\tau, N)$ keeping one of the parameters constant and varying the other.
Note that the $T$, $L$ dependence is reduced to a dependence on the single parameter $\tau$ which exclusively determines the behaviour of the system as discussed above.
Following this procedure we have computed  $\avg{\varepsilon(\tau)}$ for $N=100$,
 using both the Metropolis and the Wang-Landau algorithm.
Obviously (Fig. \ref{mesen} (a)) both algorithms give overall similar results, which can  be interpreted as  evidence for convergence. However, the Wang-Landau algorithm fails 
to describe the regime of ultra high temperatures due to the predefined limited energy space available for the simulation. The reduced specific mean potential energy $\avg{\varepsilon}$
is almost constant for low $\tau$, as the mean energy there is fully characterized by the minimum value of $\frac{V_C (\xi_1, \xi_2, \ldots, \xi_N)}{g^2}$ and 
increases almost linearly with the logarithm of the reduced temperature ($\avg{\varepsilon} \propto \log(\tau)$) for  higher temperatures. Thus for low temperatures we obtain the 
familiar $\avg{E} \sim L^{-1}$ law for the Coulomb system.  At temperatures close to $\tau_c\approx 10$ a weak change in the sign of the curvature can be observed 
(Fig. \ref{mesen} (a) (inset)), a fact  that, as we will discuss below, leads to a smooth maximum in the heat capacity $C_L$. The dependence of $\avg{E}$ on $\tau$ 
(on both $L$ and $T$) for higher temperatures can be interpreted as a consequence of the dependence of the transition temperature $T_c$  on $L$ which is confirmed by the
behaviour of $C_L$ (see below).

We have used as well the Metropolis algorithm to obtain the dependence of $\avg{\varepsilon}$ on $N$ for three  reduced temperatures, characterized by a different qualitative behaviour, 
corresponding to the low,  intermediate and high temperature regime respectively: $\tau = 10^{-2}, 10$ and $10^5$.
The diagrams of Fig. \ref{mesen} (c),(d)  show a linear relation  on a  log-log scale which indicates a power law relation between the involved
 quantities. The results of the linear fits  lead us to the following conclusions. For Fig. \ref{mesen} (b) (low temperature region) the  fit contains
not only a linear but also a logarithmic term indicating a relationship $\avg{\varepsilon} \propto N \log(N) $. This logarithmic correction is expected for
1-D Coulomb systems  in this temperature regime \cite{Dubin97}, due to the formation of a crystal and the fixed particle positions.
 For larger temperatures the relation becomes linear  resulting in a characteristic exponent  around $1.12$ for $\tau=10$ and $1.04$ for $\tau=10^5$,
 thus approaching $1$ for high temperatures ($\avg{\varepsilon} \propto N$). It is clear that since the reduced
mean specific energy $\avg{\varepsilon} $ depends on $N$ extensivity is violated, a fact that is expected for systems with long-range interactions. In summary, the mean Coulomb energy 
scales roughly as $\avg{E} \propto \frac{N^2}{L}$ which coincides with our intuition for Coulomb systems,  but looking more precisely at $T \rightarrow 0$ the scaling of the
mean potential energy takes the form (Fig. \ref{mesen} (b)):
\begin{equation}
 \avg{E}\propto \frac{N^2 \log(N)}{L}, \label{eqtl1}
\end{equation}
 which will be used below in an attempt of interpreting the thermodynamic
 limit in such a system.

\begin{figure}[htbp]
\begin{center}
\includegraphics[width=8.6cm]{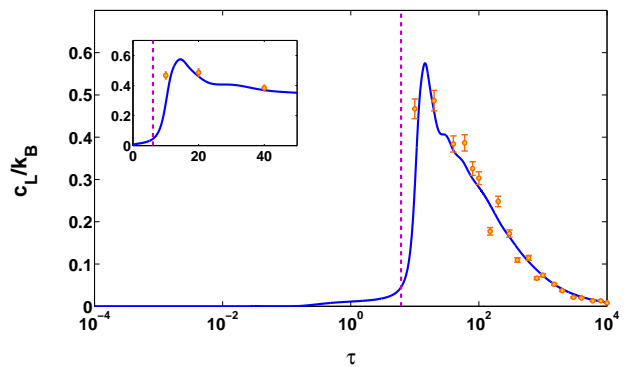}
\end{center}
\caption{\label{cvall} (color online).  (a) The temperature dependence of the specific heat capacity $\frac{c_L}{k_B}$ for $N=100$ on a semi-logarithmic scale: 
(blue line) Wang-Landau results, (orange circles) Metropolis results. The dashed purple line indicates the temperature $\tau_1$ at which the skewness becomes zero.}
\end{figure}

\textit{Specific heat capacity  $c_L$}. In order to proceed in our study of the temperature dependence of $\avg{\varepsilon}$ we have also calculated the specific heat capacity 
$c_L\equiv \frac{1}{N}\left(\frac{\partial \avg{\varepsilon}}{\partial \tau}\right)_{L}$  of the Coulomb gas confined in an one-dimensional box.
Due to the fact that numerical differentiation usually encounters large errors it is difficult to extract the heat capacity's values from the results of the Metropolis algorithm
as shown in Fig.~\ref{mesen} (a).  A calculation of the heat capacity via the energy fluctuations:
\begin{equation}
 \frac{c_L}{k_B}=\frac{1}{N}   \Lambda^2 \left(\avg{\varepsilon^2}-\avg{\varepsilon}^2\right), \label{cveq1}
\end{equation}
 proved also to be insufficient with the use of the Metropolis' results, especially in the low temperature regime. The main problem was again the large errors involved 
in the calculation of the energy's variance. When calculated with the jackknife method \cite{barkema}, they were found to have a value up to $70\%$, a fact that 
renders the results useless.
Thus, we have used instead the density of states $g(\varepsilon)$  obtained from the Wang-Landau algorithm, in order to compute the quantities $\avg{\varepsilon}, \avg{\varepsilon^2}$ 
and finally 
 the reduced specific heat capacity
via the relation (\ref{cveq1}).
We present our results for $N=100$ in Fig.~\ref{cvall} using a semi-logarithmic scale.
A smooth maximum occurs which is located at $\tau \approx 11$, thus within the temperature region in which the histogram
 of the inter-particle distance $\Delta \xi_i$ takes approximately  a Gaussian form  (Fig.~\ref{difx1} (d)) at the transition point from the crystalline to cluster phase.
 The  temperature $\tau_1$ at which the skewness of the distributions becomes zero (dashed purple line) coincides with the temperature at which the caloric
 curve starts to increase significantly. The results of Metropolis for high temperatures are found to be in accordance with the Wang-Landau results.  The complete behaviour
of $c_L$ reinforces 
the evidence that in this temperature region around $10$ there is a crossover, whose transition temperature $T_c$ decreases as $L^{-1}$ since for fixed $N=100$ we obtain
 $\tau_c \approx 10 \Rightarrow T_c \approx \frac{10 g^2}{k_B L}$.

 \textit{The thermodynamic limit.} As depicted in the dependence of the mean potential energy $\avg{E}$ on $N$, $L$ (eq. (\ref{eqtl1})) the system of ions 
confined in a 1-D box presents strong deviations from extensivity. The statistical treatment of non-extensive systems is a challenging task and many of its aspects remain still
open questions. A detailed study of the handling of non-extensivity in order to obtain thermodynamic quantities and parameters that remain finite in the  thermodynamic limit
can be found in \cite{Tsallis}. Although this procedure is beyond the scope of the present paper which aims only at the description of finite systems, we present some basic results
obtained with the methods discussed in \cite{Tsallis}. 

The main idea is to normalize the thermodynamic quantities except the entropy $S$  with the excessive powers of $N$, a fact that is justified in the context of finite systems and 
resembles the procedure we have followed to define dimensionless quantities independent of the system's size, like the reduced temperature.
Assuming a uniform distribution of charges with $\rho=g^2 \frac{N}{L}$, the mean potential energy per particle scales roughly as:
\[\frac{\avg{E}}{N}=\rho \int_{1}^{N} r^{-1} dr = \rho  \log(N)= g^2 \frac{N \log(N)}{L} \]
Note that this expression is the same as the one we have obtained for the low temperature regime (eq. (\ref{eqtl1})). The excessive dependence $N^*$ in $N$ is then given by
$N^*=N \log(N)$ and one can define  a normalized energy $\avg{E^*}\equiv \frac{\avg{E}}{N^*}$ which is a pseudo-extensive quantity \cite{Tsallis}. This is used in order to obtain the 
thermodynamic limit in the following way:
 \begin{equation}
\lim_{N\rightarrow \infty} \frac{F(T,N)}{NN^*}=\lim_{N\rightarrow \infty} \frac{\avg{E(T,N)}}{NN^*}- \lim_{N\rightarrow \infty}\frac{T}{N^*} \frac{S(T,N)}{N} \label{thl1}
\end{equation}
In order to keep entropy an extensive quantity we need to define a pseudo-intensive temperature $T^*\equiv\frac{T}{N^*}$, which leads to a normalized reduced temperature:
 \begin{equation}
\tau^*\equiv\frac{\tau}{N^*}=\frac{\tau}{N \log(N)} \label{nrt1}
\end{equation}
This quantity is dimensionless and independent of the size $(N,L)$ of the system, thus its use helps to interpret our results in the thermodynamic limit.
The critical temperature of our system is then found to be:
\begin{equation}
\tau^*_c=\frac{\tau_c}{N^*}\approx 0.022 \label{nrt2}
\end{equation}
Since $\tau_c=0.022 N \log(N)$ we conclude that indeed $\tau_c$ increases with the number of particles $N$ faster than linearly. What is more,
for $N=50$ we are led to a value $\tau_c \approx 4.2$, whereas for $N=200$ the critical value is $\tau_c \approx 23$. These values are very close to the crude results 
obtained above from the study of the mean cluster size as a function of temperature (Fig. \ref{clust2}).

\begin{figure}[htbp]
\begin{center}
\includegraphics[width=8.6cm]{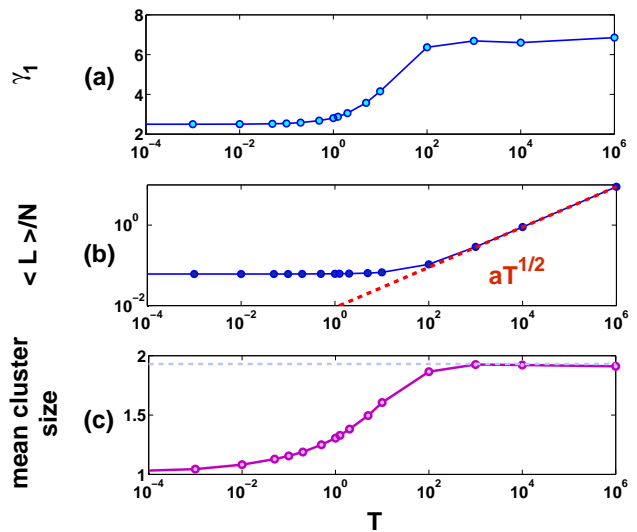}
\end{center}
\caption{\label{hosc1} (color online). Statistical properties of a system of $N=100$ ions confined in a harmonic trap with $m \omega^2=32$:
(a) The skewness of the distributions  $\rho(\Delta x_i)$ as a function of the temperature $k_B T$.
(b) The temperature dependence of the chain's mean length $\avg{L}$. (c) The mean cluster size as a function of the temperature $k_B T$. 
The horizontal dashed line indicates the value of the mean cluster size for  $N$  non-interacting 
particles confined in the same harmonic trap.}
\end{figure}

\begin{center}
 { \bf{IV. THE INFLUENCE OF THE TRAPPING POTENTIAL}} 
\end{center}
We have studied the statistical properties of a 1-D chain of ions confined in a box. Let us now briefly investigate the effect of the trapping potential on these  properties.
We focus especially on the case of the harmonic trap since it is the most frequently used and convenient trapping potential. In the low temperature regime 1-D systems of ions 
under harmonic confinement have been extensively studied \cite{Dubin97, Schiffer03, Morigi04, Morigi2004, Fishman08, Gong10}. It has been found that a crystal is obtained whose
 length  depends non trivially on the number of particles. The crystalline structure represents a non-uniform ion distribution with the opposite behaviour of  the
 one encountered for the box trap (Fig. \ref{dens1}) since the ions tend to 
accumulate in the inner region rather than the edges. This leads to a discrete but positively skewed distribution of inter-particle distances $\rho(\Delta \xi_i)$ even for ultra low temperatures 
(Fig. \ref{hosc1} (a)). For temperatures until $T\approx 1$ the skewness $\gamma_1$ is almost constant. Then it starts to increase up to $T\approx 10^2$ and from then on it
 saturates at a value close to $7$.  No minimum or zero value exists. Although the distribution  $\rho(\Delta \xi_i)$ still undergoes a transition 
from a discrete form to a continuous one with a maximum at $\Delta \xi_i$ approaching zero, it is not possible to identify a 
clear transition temperature by requiring this distribution to become symmetric. A clustering (most probable value: $\Delta \xi_i \rightarrow 0$) is observed for higher temperatures 
in the harmonic trap but the form of the distribution does not tend to an exponential, but to one with a much longer tail as indicated by the very large value of the skewness. 
 In general the deviation of the statistical behaviour of the
ions under harmonic confinement from the observed one for the box trap can to a large extent be attributed to the change of the chain's 
length with temperature (Fig. \ref{hosc1} (b)). In the low temperature regime the length of the chain is approximately constant with a value given by \cite{Dubin97}:
\begin{equation}
 L_0(N)\approx 2\left(3 N \log (N) \frac{g^2}{m \omega^2}\right)^{1/3} \label{lhos}
\end{equation}
 In our case this yields $L\approx 7$. For temperatures larger than $T\approx 10^2$ we encounter the behaviour $\avg{L}\propto T^{1/2}$, which is clearly
what would be obtained for non-interacting particles under harmonic confinement in 1-D. Thus for very large temperatures all the properties are dominated essentially by the trapping
potential.

 Concerning the formation of the clusters, sizes larger than $1$ are possible even for comparatively low temperatures (Fig. \ref{hosc1} (c)). The mean cluster size increases with 
temperature and saturates finally at  $T\approx 10^2$ at a value smaller than the respective one for the box (Fig. \ref{clust2}).  In this case too, the saturation 
value coincides  with the value of the mean cluster size obtained for an identical system in the absence of interactions and it thus constitutes a feature completely determined by 
the size of the system and the trapping potential. 
 A definition of a reduced temperature is not feasible since the Hamiltonian contains parts with different spatial dependencies. However, it is possible to use the relation obtained for the box $\tau^*=k_B T\frac{L}{N \log(N)}$
and substitute the length $L$ of the box with the minimum length  of the chain $L_0(N)$ (eq. (\ref{lhos})). We find then a critical temperature $k_B T_c\approx 1.4$. Close to this value 
the skewness and the mean length start to rise whereas the mean cluster size reaches half of its maximum value. Therefore, this could be interpreted as a critical temperature for the system of
ions under harmonic trapping.

To conclude, the transition from a crystalline to a cluster phase holds independently of the trapping potential. All the statistical properties seem to be 
affected by temperature, in a way similar to the case of the box trap and thus using the appropriate length scale, the expression obtained for the critical
 temperature (eq. (\ref{nrt2}))
provides us with valid results for different trapping potentials. However, the specific forms of the distributions $\rho(\xi_i),~ \rho(\Delta \xi_i)$ are affected by the change 
of the trapping, since the minimum configuration of the total potential $V=V_C+V_{trap}$, i.e. the shape of the crystal, is qualitatively different for each case. 
Furthermore, for high temperatures these distributions are predominantly characterized by the trapping potential, tending to the  forms that would be obtained
 in the case of non-interacting  particles confined in the same traps.

 \begin{center}
 { \bf{V.  CONCLUSIONS}} 
\end{center}

We have explored the statistical and thermodynamical properties of a system of equally charged particles confined in a box of length $L$.  This finite system exhibits 
a crossover from a crystalline to a cluster  phase  at a transition temperature $T_c$ which depends both on the number of particles $N$ and the length $L$ of the box,
as $T_c \propto \frac{N \log(N)}{L}$. This expression comes from an interpretation of the thermodynamic limit for  non extensive systems, where a normalized temperature $\tau^*$ is
introduced scaling with the system's size in the same way as the mean potential energy per particle.  The thermodynamic behaviour 
of the finite system  is properly described by the characteristic distributions and correlations of the scaled relative distances $\Delta \xi_i$ of the particles, 
whereas quantities based on the absolute positions of the particles are barely affected by the temperature. Concretely, the probability density of   $\Delta \xi_i$
undergoes a marked transition from a discrete accumulative to an exponential form as the temperature increases. In the transition regime it acquires a symmetric form
and the corresponding heat capacity shows a maximum. For high temperatures the particles form clusters whose size
grows with temperature and is finally saturating.  When the trapping conditions are altered a similar transition occurs, but the critical temperature is 
not clearly identifiable through the forms of the distributions as described above.

In order to relate the results of our simulations to concrete physical setups we need to assign specific values to the parameters. 
 Thus, for example, the temperature measured in S.I. units ($T^{SI}$) is related to the normalized temperature $\tau^*$ through: 
\begin{eqnarray}
 T^{SI}&=&\frac{g^2 N \log(N) \tau^*}{k_B L}=\frac{Z^2 e^2 N \log(N) \tau^*}{4 \pi \varepsilon_0 k_B L}\nonumber \\
&\approx& 1.7  \cdot 10^{-5} Z^2  \tau^* \frac{N \log(N)}{L}~ K m,\nonumber
\end{eqnarray}
where $Z$ is the charge of the ions. We find then from eq. (\ref{nrt2}) that $T^{SI}_c=3.7  \cdot 10^{-7} Z^2 \frac{N \log(N)}{L}~K m$. 
For cold atomic ions, the formation of quasi one-dimensional crystals with $L\approx 1~mm$ \cite{Dantan09}, $N \approx 100$ ions and e.g $Z=1$
 leads to a critical  temperature $T^{SI}_c\approx 1.7  \cdot 10^{-1}~K$, which is easily accessible experimentally. Therefore, by tuning the values of $L$ and $N$ it is 
possible to study this crystalline-clustering crossover in laboratory experiments with cold ions assuming a 1-D trapping geometry.

Let us now speculate about  potential applications of our detected crossover and related phases to high energy collisional experiments of e.g. highly charged nuclei.
 Although the geometry of the highly energetic colliding ions is quasi 2-D it is interesting to estimate some properties assuming roughly that the 1-D picture explored here
 is not a misleading description. In nuclear systems the length scales are of order $L\approx 10^{-15} m$ which for $N=100$ nucleons leads to
$T^{SI}_c\approx 10^{11}~K$ i.e. to a relative energy $E_c\approx 10~MeV$. The nuclear mean binding energy is of the order of $8 MeV$ per nucleon resulting in a
 total energy of $800MeV$ for the nucleus, somewhat larger than $E_c$. Thus the nucleons confined in the colliding nuclei  could be in the cluster phase. 
Finally, we remark that the crossover physics analyzed in this work could also appear in one-dimensional systems showing other long-range interactions,
 such as dipolar chains \cite{Armstrong12}.

 \begin{center}
 { \bf{APPENDIX}} 
\end{center}
Here we present  a proof that the global minimum energy configuration is the only minimum that the potential of the finite 1-D ion chain can acquire.
 We want only permutationally different configurations so we assume that we have $N+2$ particles that are ordered in 
the sense that $\xi_0< \xi_1 < \ldots < \xi_N < \xi_{N+1}$ ($\xi_0$ is the position of the first particle and $\xi_{N+1}$ the position of
the last one). We have shown (Sec. IIIA) that in order to obtain a minimum the first particle has to be fixed at the left edge of the segment, i.e. $\xi_0=0$,
whereas the last particle should always occupy its right edge $\xi_{N+1}=1$ (the length of box is $L=1$).
Thus, the potential is given by the expression:
\begin{equation}
V = \sum_{i=0}^{N+1}\sum_{j=0}^{i-1} \frac{1}{\xi_i-\xi_j} =\sum_{i=1}^{N}\sum_{j=1}^{i-1} \frac{1}{\xi_i-\xi_j}+\sum_{i=1}^{N}\left(\frac{1}{\xi_i}+\frac{1}{1-\xi_i}\right) \nonumber
\end{equation}
Differentiating this expression with respect to $\xi_k$ we obtain:

\begin{equation}
\pd{V}{\xi_k} =-\sum_{i=1}^{k-1} \frac{1}{(\xi_k-\xi_i)^2} +\sum_{i=k+1}^{N} \frac{1}{(\xi_i-\xi_k)^2}-\frac{1}{\xi_k^2}+\frac{1}{(1-\xi_k)^2} \nonumber
\end{equation}

A second differentiation with respect to $\xi_l$ leads to:
\begin{eqnarray}
\frac{\partial^2 V}{\partial \xi_l \partial \xi_k} &=&  \sum_{i=1}^{k-1} \frac{2}{(\xi_k-\xi_i)^3} \delta_{kl}- \sum_{i=1}^{k-1} \frac{2}{(\xi_k-\xi_i)^3} \delta_{il} \nonumber \\
&-& \sum_{i=k+1}^{N} \frac{2}{(\xi_i-\xi_k)^3} \delta_{il}+  \sum_{i=k+1}^{N} \frac{2}{(\xi_i-\xi_k)^3} \delta_{kl} \nonumber \\
&+&\left(\frac{2}{\xi_k^3}+\frac{2}{(1-\xi_k)^3}\right) \delta_{kl} \nonumber
\end{eqnarray}

Therefore,  we obtain for the Hessian matrix $H$ of our potential  $V$ ($H_{ij}\equiv \frac{\partial^2 V}{\partial \xi_i \partial \xi_j}$):

\begin{equation}
H_{kk}=\frac{\partial^2 V}{\partial \xi_k \partial \xi_k} = \frac{2}{\xi_k^3}+\frac{2}{(1-\xi_k)^3}+2 \sum_{i\neq k} \frac{1}{\abs{\xi_i-\xi_k}^3}>0 \nonumber
\end{equation}
whereas the off diagonal elements ($l\neq k$) are given by:
\begin{equation}
H_{lk}=\frac{\partial^2 V}{\partial \xi_l \partial \xi_k} = - \frac{2}{\abs{\xi_k-\xi_l}^3}<0 \nonumber
\end{equation}

It is known from multivariate calculus that if for arbitrary $d\vec{\xi}$, $\vec{\xi}$ it holds that:
\begin{equation}
d^2 V= 2 \sum_{i=1}^N \sum_{j=1}^N H_{ij}(\vec{\xi}) d\xi_i d\xi_j  >0 \nonumber
\end{equation}
then the function $V$ is strictly convex and it possesses at most one minimum.
In our case:
\begin{eqnarray}
d^2 V &=& \sum_{k=1}^N H_{kk} d^2 \xi_k + \sum_{k=1}^N \sum_{\substack{l=1\\   l \neq k}}^N H_{lk} d\xi_k d\xi_l \nonumber \\
&=& 2 \sum_{k=1}^N \left(\frac{1}{\xi_k^3}+\frac{1}{(1-\xi_k)^3}\right)d^2 \xi_k\nonumber \\ 
&+&\sum_{k=1}^N\sum_{\substack{l=1\\   l \neq k}}^N \frac{1}{\abs{\xi_k-\xi_l}^3}\left(d\xi_k-d\xi_l\right)^2 \nonumber
\end{eqnarray}

It is obvious that $\sum_{k=1}^N\sum_{\substack{l=1\\   l \neq k}}^N \frac{1}{\abs{\xi_k-\xi_l}^3}\left(d\xi_k-d\xi_l\right)^2 \geq 0$ and 
$\sum_{k=1}^N \left(\frac{1}{\xi_k^3}+\frac{1}{(1-\xi_k)^3}\right)d^2 \xi_k>0$ for $d\vec{\xi} \neq 0$, so the quantity $d^2 V$ is always positive
and the potential a convex function with only one  minimum. 

In general if a trapping potential of  the form $V_{trap}=A \sum_{k=1} \xi_k^n$, with $A>0$ and  an even number $n \geq 2 $ is chosen,
 it is evident from the above procedure that for the total potential $V=V_C+V_{trap}$ it holds:
\begin{equation}
d^2 V= n (n-1) A  \sum_{k=1}^N \xi_k^{n-2} d^2 \xi_k +\sum_{k=1}^N\sum_{\substack{l=1\\   l \neq k}}^N \frac{1}{\abs{\xi_k-\xi_l}^3}\left(d\xi_k-d\xi_l\right)^2, \nonumber 
\end{equation}
which is also positive definite and thus the total potential possesses again  a single minimum.

\begin{center}
\begin{acknowledgments}
We thank  B. Liebchen for stimulating discussions. 
\newline
 A.  Z. thanks the International Max Planck Research School for Ultrafast Imaging and  Structural Dynamics for a PhD scholarship.
\end{acknowledgments}
\end{center}

\end{document}